\documentclass[reprint,nofootinbib,aps,prd,table]{revtex4-2}

\usepackage{graphicx}
\usepackage{dcolumn}
\usepackage{booktabs}
\usepackage{bm}
\usepackage{hyperref}
\usepackage{autobreak}
\usepackage{orcidlink}
\usepackage{physics}
\usepackage{amsmath, amssymb} 
\usepackage{mathtools}
\usepackage{ragged2e}
\usepackage{soul} 

\usepackage{multirow}
\usepackage{array}
\newcolumntype{P}[1]{>{\centering}p{#1}}

\renewcommand{\arraystretch}{1.5}

\usepackage{siunitx}
\DeclareSIUnit \parsec {pc}
\newcommand{\ds}{\displaystyle}
\usepackage{tikz,lipsum,lmodern}
\usepackage[T1]{fontenc}
\usepackage[most]{tcolorbox}
\usepackage{wrapfig}

\def\aj{AJ}

\tcbset{colback=green!5,colframe=green!35!black,fonttitle=\bfseries}

\definecolor{deltared}{RGB}{255,220,220}
\definecolor{deltablue}{RGB}{220,225,255}

\newcommand{\lcdm}{$\Lambda$CDM} 
 
\newcommand{\wowacdm}{$w_0w_a$CDM} 
\newcommand{\wowa}{($w_0, w_a$)} 
\newcommand{\Okcdm}{$\Omega_k$CDM} 
\newcommand{\Om}{\Omega_\mathrm{m}}
\newcommand{\Ocdm}{\Omega_\mathrm{cdm}}
\newcommand{\Ok}{\Omega_\mathrm{K}}
\newcommand{\Or}{\Omega_\mathrm{R}}

\newcommand{\Ob}{\Omega_\mathrm{b}}
\newcommand{\Obhtwo}{\Omega_\mathrm{b}h^2}
\newcommand{\Ocdmhtwo}{\Omega_\mathrm{cdm}h^2}

\newcommand{\DM}{D_\mathrm{M}}
\renewcommand{\DH}{D_\mathrm{H}}

\newcommand{\dchisq}{\Delta \chi^2}
\newcommand{\plcdm}{\mathbf{p}^{\Lambda \rm CDM}}
\newcommand{\ME}{M^{\rm E}}

\newcommand{\zt}{z_{\rm t}}
\newcommand{\ztM}{z_{\rm tM}}
\newcommand{\ztH}{z_{\rm tH}}
\newcommand{\HnodE}{H_0^{\rm E}}

\newcommand{\kmsMpc}{\,{\rm km\, s^{-1}Mpc^{-1}}}
\newcommand{\kms}{\,{\rm km\, s^{-1}}}
\newcommand{\Mpc}{\,{\rm Mpc}}

\begin{document}
\preprint{000-000-000}

\title{On the Difficulties with Late-Time Solutions for the Hubble Tension}

\author{Prakhar Bansal\,\orcidlink{0009-0000-7309-4341}}
 \email{prakharb@umich.edu}

\author{Dragan Huterer\,\orcidlink{0000-0001-6558-0112}}%
 \email{huterer@umich.edu}
\affiliation{
Department of Physics and Leinweber Institute for Theoretical Physics, University of Michigan, 450 Church St, Ann Arbor, MI 48109
}

\date{\today}

\begin{abstract}
We explore the notion that cosmological models that modify the late-time expansion history cannot simultaneously fit the SH0ES collaboration's measurements of the Hubble constant, DESI baryon acoustic oscillations data, and Type Ia supernova distances. Adopting a few simple phenomenological models, we quantitatively demonstrate that a satisfactory fit with a model with late-time expansion history  can only be achieved if one of the following is true: 1) there is a sharp step in the absolute magnitude of Type Ia supernovae at very low redshift, $z\sim 0.01$, or 2)  the distance duality relation, $d_L(z)=(1+z)^2d_A(z)$, is broken. Both solutions are trivial in that they effectively decouple the calibrated SNIa measurements from other data, and this qualitatively agrees with previous work built on studying specific dark-energy models. We also identify a less effective class of late-time solutions with a transition at $z\simeq 0.15$ that lead to a more modest improvement in fit to the data than models with a very low-z transition. Our conclusions are largely unchanged when we include surface brightness fluctuation distance measurements, with their current systematic uncertainties, to our analysis. We finally illustrate our findings by studying a  physical model which, when equipped with the ability to smoothly change the absolute magnitude of Type Ia supernovae, partially resolves the Hubble tension. 
\end{abstract}

\maketitle


\section{Introduction}
Recent evidence for the so-called "Hubble tension" -- a discrepancy in the measurements of the Hubble constant between low-redshift direct determinations and higher-redshift inferences -- has caused waves in cosmology and emerged as a forefront challenge to the standard cosmological model with dark matter and dark energy (\lcdm). Direct measurements provided by the SH0ES collaboration \cite{Riess:2021jrx} indicate that $H_0=73.04\pm 1.04\kmsMpc$. In contrast, parameter inference from the CMB gives $H_0=67.14\pm 0.47\kmsMpc$ \cite{Planck:2018}, while the combination of baryon acoustic oscillation (BAO) measurements from the Data Release 2 of the Dark Energy Spectroscopic Instrument (DESI) and the big bang nucleosynthesis prior gives $H_0=68.5\pm 0.51\kmsMpc$ \cite{DESI:2025zgx}. Consequently, measurements by the SH0ES collaboration are 5.2$\sigma$ and 3.9$\sigma$ apart from the CMB and BAO+BBN measurements, respectively \footnote{The latest measurement of $H_0$ from the H0DN collaboration\cite{H0DN:2025lyy} raises the Hubble Tension to a 7.1$\sigma$ level}. While there exist multiple other methods for local (at redshift $z\ll 1$) and global ($z\simeq 1$ or higher) determination of the Hubble constant, these methods are currently not as mature as the aforementioned measurements that are in tension; see \cite{Verde:2019ivm,Tully:2023bmr} for reviews. 

Exploration of possible systematic errors in these measurements and their inability to account for the Hubble tension has caused major effort on the theory front.  It is fair to say that no compelling theoretical solution to the Hubble tension has yet been identified. Roughly, these theoretical solutions come in two flavors: those that change the early-universe physics, typically changing the sound horizon, and those that affect the late-time expansion history (for comprehensive reviews, see \cite{DiValentino:2021izs,Schoneberg:2021qvd,Abdalla:2022yfr}). Both kinds of explanations increase the $H_0$ inferred from the CMB and BAO to better agree with the SH0ES value. 

The difficulty of explaining Hubble tension with theoretical models is well known \cite{DiValentino:2021izs}. Early-time dark-energy solutions to this problem, reviewed in \citet{Poulin:2023lkg}, technically work but are very fine tuned. The situation is even more dire with \textit{late}-time solutions to Hubble tension; it has been pointed out that they are essentially not feasible \cite{Poulin:2018zxs,Aylor:2018drw,DiValentino:2019exe,Knox:2019rjx,Dhawan:2020xmp,Cai:2021weh,Efstathiou:2024xcq,Poulin:2024ken,Pedrotti:2025ccw} except in extremely fine-tuned scenarios \cite{Benevento:2020fev}.

The stakes were raised recently with the finding by DESI and other data finding some hints for dynamical dark energy. Adopting the popular parameterization of the equation of state of dark energy $w(a) = w_0 + w_a(1-a)$ \cite{Linder:2002et,Chevallier:2000qy}, where $w_0$ and $w_a$ are free parameters and $a$ is the scale factor, the analysis of DESI data, combined with cosmic microwave background (CMB) and Type Ia supernovae (SNIa) favor values with $w_0 > -1$ and $w_a < 0$, and depart from the standard cosmological model with vacuum energy ($w_0=-1, w_a=0$) at the statistical level between 2.8$\sigma$ and 4.2$\sigma$ \cite{DESI:2024mwx,DESI:2024kob,DESI:2024aqx,DESI:2025zgx}. This finding, which was followed up widely in the literature (e.g.\ \cite{Cortes:2024lgw, Berghaus:2024kra,Mukherjee:2024ryz,Chan-GyungPark:2024mlx,Dinda:2024ktd,Reboucas:2024smm,Patel:2024odo,Chudaykin:2024gol,Jiang:2024xnu,Wolf:2024eph,Park:2024pew,Giare:2024oil,Simon:2024jmu,Linder:2024rdj,Lewis:2024cqj,Reeves:2025axp,Shajib:2025tpd,DESI:2025fii,Giare:2025pzu,Pan:2025psn,Bansal:2025ipo,Jedamzik:2025cax,Lee:2025yah,Poulin:2025nfb,Cai:2025mas,Herold:2025hkb,Keeley:2025rlg,Li:2025ops,Ishak:2024jhs,Toomey:2025xyo}) may be related to the Hubble tension, though at present it appears to be separate from it: for example, in the \wowacdm\ model, the derived value of the Hubble constant is even farther away from the SH0ES measurement than in \lcdm\ \cite{DESI:2025zgx}.

Our goal in this work is to explain as clearly as possible, and illustrate with simple examples, the difficulty that models that modify late-time expansion history of the universe face to explain  the combination of SH0ES measurements of the Hubble constant, data from Type Ia supernovae, and baryon acoustic oscillation measurements from the DESI survey. In order to do so, we do not present a comprehensive scan through classes of dark-energy models, but rather investigate a few parameterized models for the expansion history that, in our view, encompass a range of physical possibilities. We also illustrate results with a physical, albeit fine-tuned, model. 

Our paper is organized as follows. In Sec.~\ref{sec:method}, we describe the methodology and our models. In Sec.~\ref{sec:results}, we present our results, while in Sec.~\ref{sec:Wolf_model} we give a worked physical example of a model that modifies late-time expansion history in a rather drastic way, yet still does not have a sufficiently good improvement in the fit. We conclude in Sec.~\ref{sec:concl}.


\begin{figure*}[t]
\includegraphics[width=0.9\linewidth]{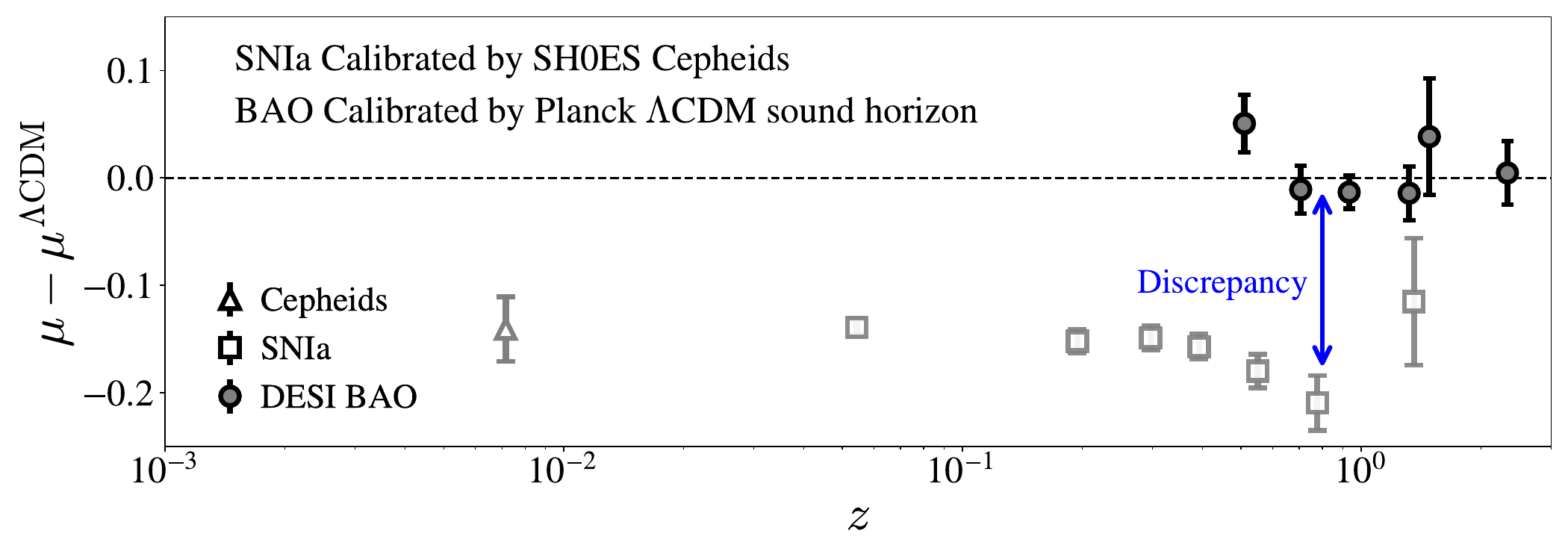}
\includegraphics[width=0.9\linewidth]{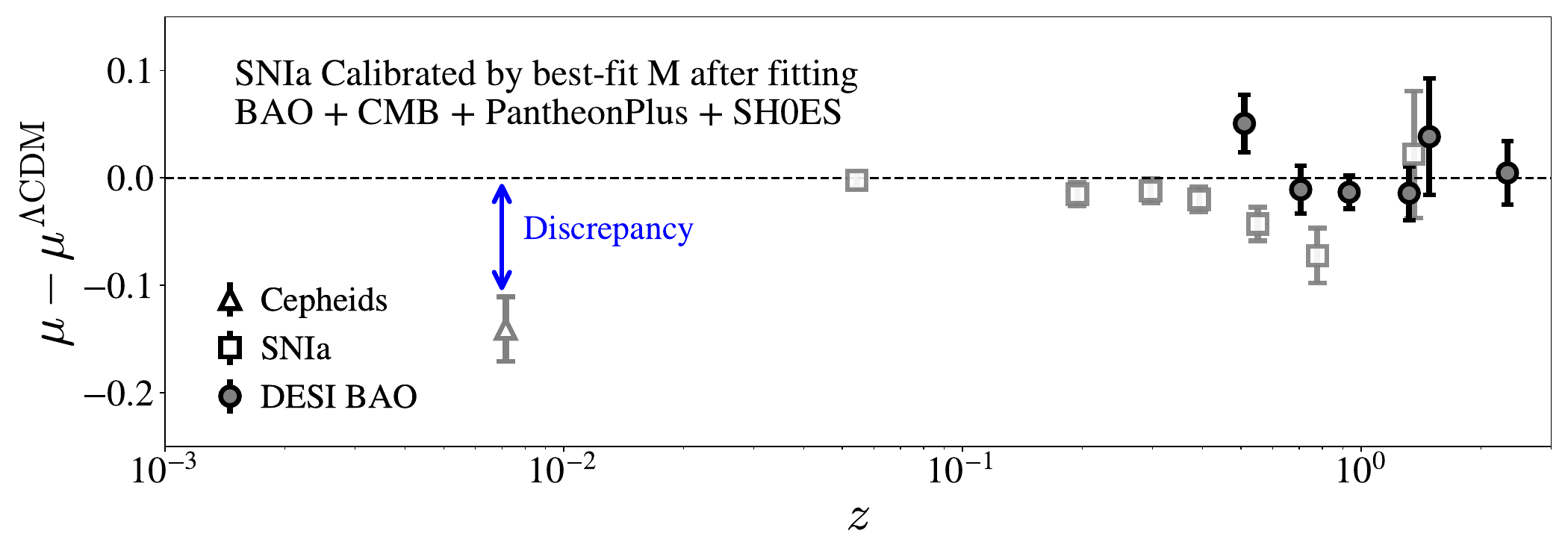}

    \caption{Illustration of the inability of low-redshift dark-energy models to simultaneously fit the SH0ES measurements and other cosmological data. In both panels, we show the distance modulus relative to one predicted in the best-fit \lcdm\ model as a function of redshift, along with the Cepheid measurements by SH0ES collaboration, SNIa data from DESY5, and BAO data from DESI DR2. The top panel shows the scenario in which SNIa are anchored to the SH0ES measurement; then a smooth model is unable to simultaneously fit the SNIa and BAO data in the region where they overlap. The lower panel shows the scenario where the SNIa data are instead anchored to BAO distances (as actually preferred by a global cosmological fit);  then the Cepheid measurement is inconsistent with the rest of the Hubble diagram in which it effectively introduces a sharp break.  See text for more details.}
    \label{fig:tension}
\end{figure*}

\section{Methodology, Models and Data}
\label{sec:method}

We assume a spatially flat cosmology and the validity of general theory of relativity throughout. The transverse and line-of-slight comoving distances are respectively
\begin{equation}
    \DM(z) = c\int_{0}^{z}  \frac{dz^\prime}    {H(z^\prime)}; 
    \quad 
    \DH(z) = \frac{c}{H(z)}~,
    \label{eq:DM_DH}
\end{equation}
where $c$ is the speed of light.  Here the expansion rate can be written in terms of the Hubble constant as $H(z)=H_0E(z)$. We now describe the cosmological models that we consider in this paper. 

\vspace{0.5cm}
\noindent
\ul{\textbf{\lcdm\, model}}: In this model the expansion history $E(z)$ takes on the standard expression  
\begin{align}
E(z) &= \sqrt{\Om(1+z)^3+\Or(1+z)^4+(1-\Om-\Or)}\notag\\[0.2cm]
\Om &\equiv \ds\left (\frac{100}{H_0}\right )^2 \left (\Ocdmhtwo+\Obhtwo\right )\,,
    \label{eq:Ez}
\end{align}
where $\Om, \Or, \Ocdm$, and $\Ob$ are respectively energy densities of matter, radiation, cold dark matter and baryons relative to critical, $H_0$ is the Hubble constant, and $h=H_0/(100\kmsMpc)$. The SNIa apparent magnitude is given by another standard expression (to be modified in the next two models we introduce further below)
\begin{equation}
m(z) = \mu(z)+M
\end{equation}
where $\mu(z)$ is the distance modulus defined as 
\begin{equation}
    \mu(z) = 5\log\left (\frac{d_L(z)}{1\Mpc}\right )+25
\end{equation} 
where $d_L(z) = (1+z)D_M(z)$ is the luminosity distance and $z$ is the redshift. The redshift used in the prefactor here is the helicoentric redshift $z_{\text{hel}}$ from the PantheonPlus catalog. In this model, the parameters are
\begin{equation}
    \mathbf{p}^{\Lambda {\rm CDM}} \in \{\Ocdmhtwo, \Obhtwo, H_0\}.
    \label{eq:p_lcdm} 
\end{equation}

We next consider extended models with more freedom than \lcdm\, which aim to better fit the combined cosmological data that include SH0ES. 

We have assumed a flat cosmological model thus far. As an alternative to flat \lcdm, in our analysis we also briefly consider a curved \lcdm, with one additional parameter, $\Ok$, and with the full parameter set
\begin{equation}
    \mathbf{p}^{\Ok {\rm CDM}} \in \{\plcdm,  \Ok\}.
    \label{eq:p_okcdm} 
\end{equation}

\vspace{0.5cm}
\noindent
\ul{\textbf{Hstep model}}: This is the simplest model, in which we allow for a transition in the Hubble parameter at redshift $\zt$. Mathematically, the expansion-rate step transition is defined via 
\begin{equation}
H(z) = 
\begin{cases}
H_0 E(z), & \text{if } z \leq \zt \\[0.1cm]
\HnodE E(z), & \text{if } z > \zt
\end{cases}
\label{eq:Hstep}
\end{equation}
where $\HnodE$ is a new parameter in general different from the late-time Hubble constant $H_0$. In this model, the matter density corresponds to the the early-time physical density in mass, that is
\begin{equation}
    \Om \equiv \ds\left (\frac{100}{\HnodE}\right )^2 \left (\Ocdmhtwo+\Obhtwo\right ).
\end{equation}
The parameters in this model are 
\begin{equation}
    \mathbf{p}^{\rm Hstep} \in \{\plcdm, \HnodE, \zt \}.
\end{equation}
where $\plcdm$ is the parameter set in \lcdm\ listed in Eq.~(\ref{eq:p_lcdm}). There are therefore two additional parameters in this model relative to the standard cosmological constant plus cold-dark-matter scenario.

\vspace{0.5cm}
\noindent
\ul{\textbf{Mstep model}}: In this model, we allow for a sharp change in the absolute magnitude of SNIa at the transition redshift $z_t$, but keep all other properties as in \lcdm. The new parameters in the model are the magnitude change and the redshift at which it occurs $\zt$; the apparent magnitude is now given by
\begin{equation}
m(z) = 
\begin{cases}
\mu(z)+M, & \text{if } z \leq \zt \\[0.1cm]
\mu(z)+\ME, & \text{if } z > \zt
\end{cases}
\label{eq:Mstep}
\end{equation}
where $\ME$ is a new parameter, different in general from the late-time absolute magnitude $M$, and $\zt$ is the transition redshift. The parameter set is then
\begin{equation}
    \mathbf{p}^{\rm Mstep} \in \{\plcdm, \ME, \zt\},
\end{equation}
Similar to the Hstep model, the Mstep model also contains two additional parameters relative to \lcdm.

\vspace{0.5cm}
\noindent
\ul{\textbf{Hstep+Mstep model}}: Here, we allow for both a transition in the absolute magnitude of SNIa at the transition redshift $\ztM$ and a transition in the Hubble parameter at redshift $\ztH$. The $M$ transition is described in Eq.~(\ref{eq:Mstep}), while the $H$ transition is defined as in Eq.~(\ref{eq:Hstep}).
The parameters in this model are 
\begin{equation}
    \mathbf{p}^{\rm Hstep+Mstep} \in \{\plcdm, \ME, \HnodE, \ztM, \ztH \}. 
\end{equation}
There are therefore four additional parameters in this model relative to \lcdm.

\vspace{0.5cm}
\noindent
\ul{\textbf{$w_0w_a$+Mstep model}}: Finally, we use the more flexible \wowa\ parametrization \cite{Linder:2002et} to model the expansion history and similar to previous models, allow for a transition in the absolute magnitude of SNIa at the transition redshift $z_t$. The $M$ transition is same as described in Eq.~(\ref{eq:Mstep}), while the Hubble parameter in the \wowa\ model is given by 

\begin{align}
H(z) &= H_{0}\Big[\,\Omega_{m}(1+z)^{3} + \Omega_{R}(1+z)^{4}\;+\\
& (1-\Omega_{m}- \Omega_{R})(1+z)^{3(1+w_{0}+w_{a})}
\exp\!\left(-\frac{3w_{a}z}{1+z}\right)\,\Big]^{1/2}.\notag
\end{align}

The parameters of this model are 
\begin{equation}
    \mathbf{p}^{w_0w_a \rm +Mstep} \in \{\plcdm, w_0, w_a, \ME, \zt \}.
\end{equation}
Similar to Hstep+Mstep model, there are four additional parameters in this model relative to \lcdm.

\vspace{0.5cm}
To carry out our analysis, we modify the  cosmological code \texttt{CAMB} \cite{Lewis:1999bs} to compute the background quantities like Hubble rate, distances and supernova magnitudes in our modified-H model. To obtain the constraints on the cosmological parameters of our model we use Monte Carlo Markov Chain (MCMC) sampler in \texttt{Cobaya} \cite{Torrado:2020dgo}. For our MCMC chains, we use the default convergence criteria of Gelman and Rubin R statistic $<$  0.01. To calculate the means, confidence intervals and likelihood distributions for our model parameters, we use \texttt{GetDist} \cite{Lewis:2019xzd} code with our converged MCMC chains.\\

We make use the following data sets:\\

\textbf{DESI DR2 BAO.}
We use the measurements from the DESI BAO analysis from its second data release (henceforth DESI DR2 BAO, or just DESI), and adopt the 12 distance measurements, and their covariance, as quoted in Table I of \cite{DESI:2024mwx} and validated in supporting DESI DR2 publications \cite{DESI:2024uvr,DESI:2024lzq}. To make our analysis simple and as model-independent as possible, we do not use the additional information from the full-shape clustering of DESI sources \cite{DESI:2024hhd}.\\

\textbf{Compressed CMB data.} As explained in \cite{Bansal:2025ipo}, for the class of models explored in this paper, it is not possible to use the full CMB power-spectrum-based likelihood. Hence, we rely on the compressed CMB likelihood \cite{Bansal:2025ipo} which involves the ``shift" parameter $R$, the angular location $\ell_a$ as well as the physical baryon density $\Obhtwo$\\

\textbf{Calibrated SNIa:} We make use of the full PantheonPlus+SH0ES catalog, including the calibrator sample, following the procedure adopted in the official analysis~\citep{Brout:2022vxf, Scolnic:2021amr, Riess:2021jrx}. The PantheonPlus compilation consists of distance modulus measurements from 1701 light curves of 1550 spectroscopically confirmed Type~Ia supernovae spanning the redshift range $0.001 < z < 2.26$. The low-redshift subset is calibrated using 77 nearby SNe~Ia hosted in galaxies with Cepheid distance measurements, forming the SH0ES distance ladder that anchors the absolute magnitude of SNe~Ia. Unlike the commonly used approach of applying a prior on the absolute magnitude $M$ to incorporate the SH0ES calibration, we instead include the full set of calibrators directly in the analysis, thereby accounting for the $\Delta \chi^2$ contributions self-consistently. Additional discussion of the subtleties associated with using a magnitude prior for the SH0ES calibration is provided in Appendix~\ref{app:Mprior}.

\section{Trouble brewing in the Hubble diagram: a preview}\label{sec:preview}

We illustrate in Fig.~\ref{fig:tension} the fundamental difficulty of simultaneously fitting SH0ES data and the SNIa and BAO data. In both panels, we show the distance modulus ($\mu$) relative to one predicted in the best-fit need to specify which best-fit \lcdm\ model ($\mu^{\Lambda {\rm CDM}}$) as a function of redshift. We also show the Cepheid measurements by SH0ES collaboration \cite{Riess:2021jrx}, SNIa data from DESY5 \cite{DES:2024jxu}, and BAO data from DESI DR2 \cite{DESI:2025zgx}. 

The top panel of Fig.~\ref{fig:tension} shows the scenario in which SNIa are anchored to the SH0ES measurement; in this case, a smooth model is unable to simultaneously fit the SNIa and BAO data in the region where they overlap. It would take a low-redshift model with \textit{extremely} rapid oscillations in redshift -- with variations on a scale of $\Delta z\simeq 0.01$ or faster --  to fit both SNIa and BAO data, something that does not seem physically plausible. 

The lower panel of Fig.~\ref{fig:tension} shows the scenario where the SNIa data are instead anchored to BAO distances, and  there absolute magnitude matches that inferred from the BAO (this is sometimes referred o as the ``inverse distance ladder" \cite{BOSS:2014hhw}). This situation is actually preferred by a global cosmological fit to SH0ES+BAO+SNIa, as the BAO data provide a stronger ``pull" on SNIa  than the SH0ES data point. In this case, it is immediately transparent that the Cepheid measurement becomes inconsistent with the rest of the Hubble diagram in which it effectively introduces a sharp break.  Specifically, SNIa now favor the absolute magnitude of $M_{\rm SNIa}=-19.40 \pm 0.0092$ , while the SH0ES measurement is $M_{\rm SH0ES}=-19.25\pm 0.027$.

Overall, Fig.~\ref{fig:tension} illustrates the steep challenge faced by models that modify the low-redshift expansion history. Note, in particular, that even an arbitrarily sharp transition in the expansion rate at low-redshift does not alleviate the problem, because it is a transition in the absolute magnitude $M$ that is apparently required by the combined SH0ES and SNIa/BAO data \cite{Knox:2019rjx,Efstathiou:2024xcq,Poulin:2024ken}.

\begin{table*}[t]
\centering
\renewcommand{\arraystretch}{1.2}
\setlength{\tabcolsep}{6pt}
\begin{tabular}{l  P{1.4cm}c c c c c c}
\midrule
\multicolumn{8}{c}{\textbf{Poor fit to DR2 BAO + CMB + PantheonPlus + SH0ES data}} \\
\midrule
\textbf{Model} & $\boldsymbol{\chi^2_{\mathrm{total}}}$ &\cellcolor{deltablue} $\boldsymbol{\Delta\chi^2}$& $\boldsymbol{\chi^2_{\mathrm{BAO+CMB}}}$ & $\boldsymbol{\chi^2_{\mathrm{Panth+SH0ES}}}$ & $\ME$ & $M$ & $\Delta$DOF \\
\midrule
\addlinespace
\addlinespace
\textbf{LCDM} & 1500 & \cellcolor{deltablue}0 & 17 & 1482 & --- & $-19.399$ &0  \\
\addlinespace
\textbf{LCDM} (PantheonPlus + SH0ES only) & -- &\cellcolor{deltablue} -- & -- & 1452  & --- & $-19.243$ & 0 \\
\addlinespace
\textbf{\Okcdm} & 1490 & \cellcolor{deltablue}$-10$ &12& 1478 & --- & $-19.384$ & 1  \\
\addlinespace
\textbf{\wowacdm} & 1495 &\cellcolor{deltablue} $-5$ & 15 & 1480  & --- &  $-19.377$&  2\\
\addlinespace
\textbf{HStep} & 1498 & \cellcolor{deltablue}$-2$ &19 & 1479 & $-19.395$ & $-19.395$ & 2 \\
\midrule
\addlinespace
\multicolumn{8}{c}{\textbf{Good fit to DR2 BAO + CMB + PantheonPlus + SH0ES data}} \\
\midrule
\textbf{Model} & $\boldsymbol{\chi^2_{\mathrm{total}}}$ & \cellcolor{deltablue}$\boldsymbol{\Delta\chi^2}$ & $\boldsymbol{\chi^2_{\mathrm{BAO+CMB}}}$ & $\boldsymbol{\chi^2_{\mathrm{Panth+SH0ES}}}$ & $\ME$ & $M$ & $\Delta$DOF \\
\midrule
\addlinespace
\addlinespace
\textbf{Mstep} & 1460 &\cellcolor{deltablue} $-40$ & 17 & 1443 & $-19.414$ & $-19.246$ & 2 \\
\addlinespace
\textbf{Hstep+Mstep}  ($\ztH=\ztM$) & 1458  &\cellcolor{deltablue}$-42$ & 18 & 1440 & $-19.419$ & $-19.250$ & 3 \\
\addlinespace
\textbf{Hstep+Mstep} & 1457 &\cellcolor{deltablue} $-43$ & 17 & 1440 & $-19.423$ & $-19.247$ & 4 \\
\addlinespace
\textbf{$w_0w_a$+Mstep} & 1457 &\cellcolor{deltablue} $-43$ &15 & 1443 & $-19.408$ & $-19.250$ & 4 \\
\bottomrule
\end{tabular}
\caption{Goodness of fit values for various dataset combinations using DR2 BAO, CMB, and full SNIa data comprised of PantheonPlus and SH0ES. The total $\chi^2$ is shown along with $\Delta\chi^2$ relative to the best-fit \lcdm\ model, and this is followed by $\chi^2$ contributions that come from BAO+CMB data only, and from PantheonPlus+SH0ES data only. We also show the best-fits for the absolute magnitude $M$ and, in models with a transition in absolute magnitude, we also show the early-time (before-transition) value, $\ME$. The last column shows the number of degrees of freedom of a given model relative to \lcdm. The top section of the table shows models that show a very small improvement of fit relative to \lcdm, while the bottom section shows those with a substantially improved fit. }
\label{tab:results}
\end{table*}

\begin{figure*}[t]
\includegraphics[width=0.9\linewidth]{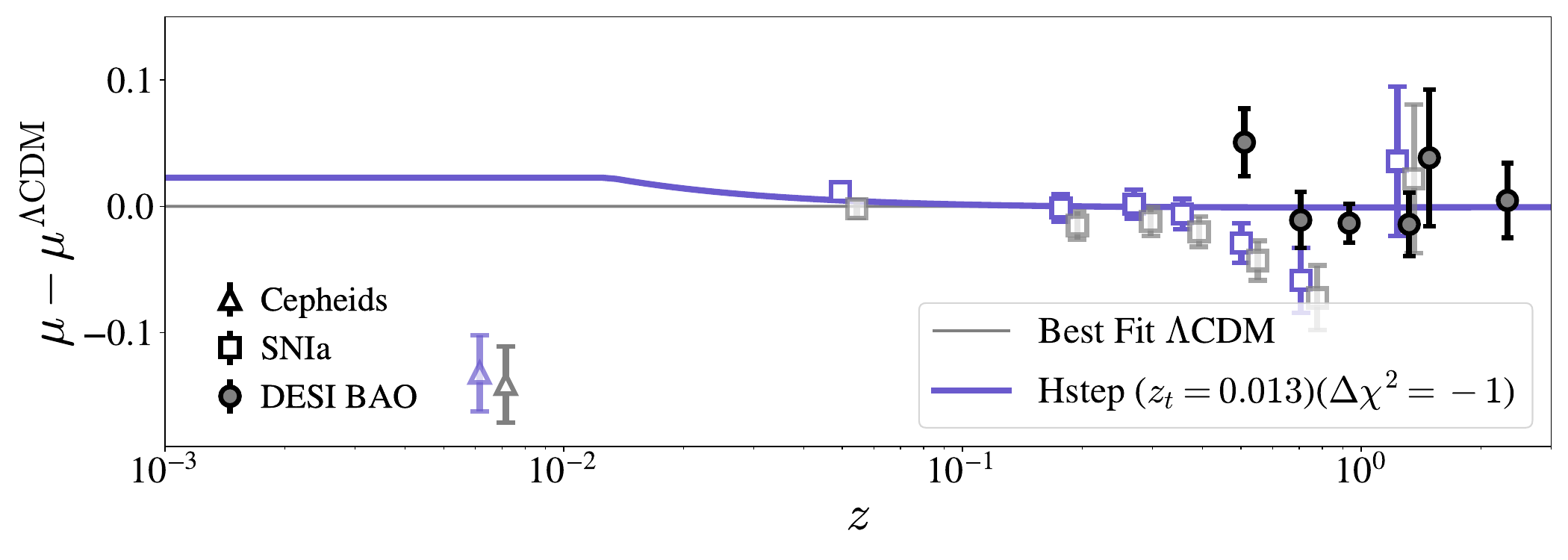}
\includegraphics[width=0.9\linewidth]{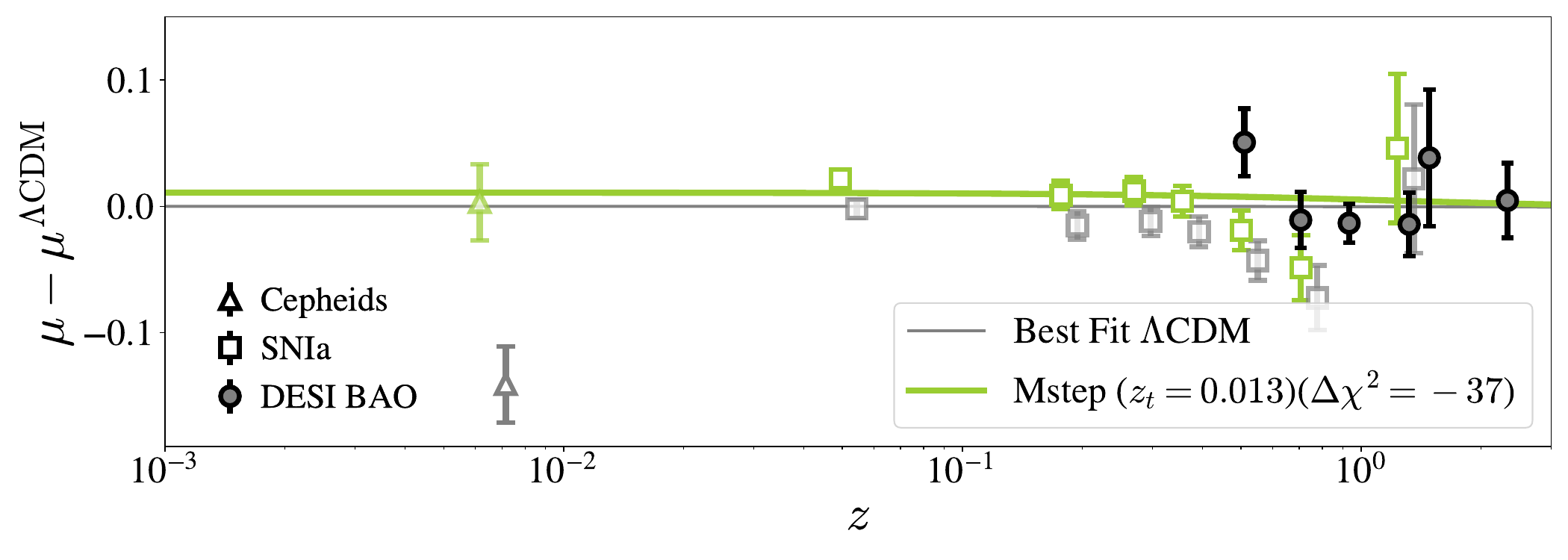}
\includegraphics[width=0.9\linewidth]{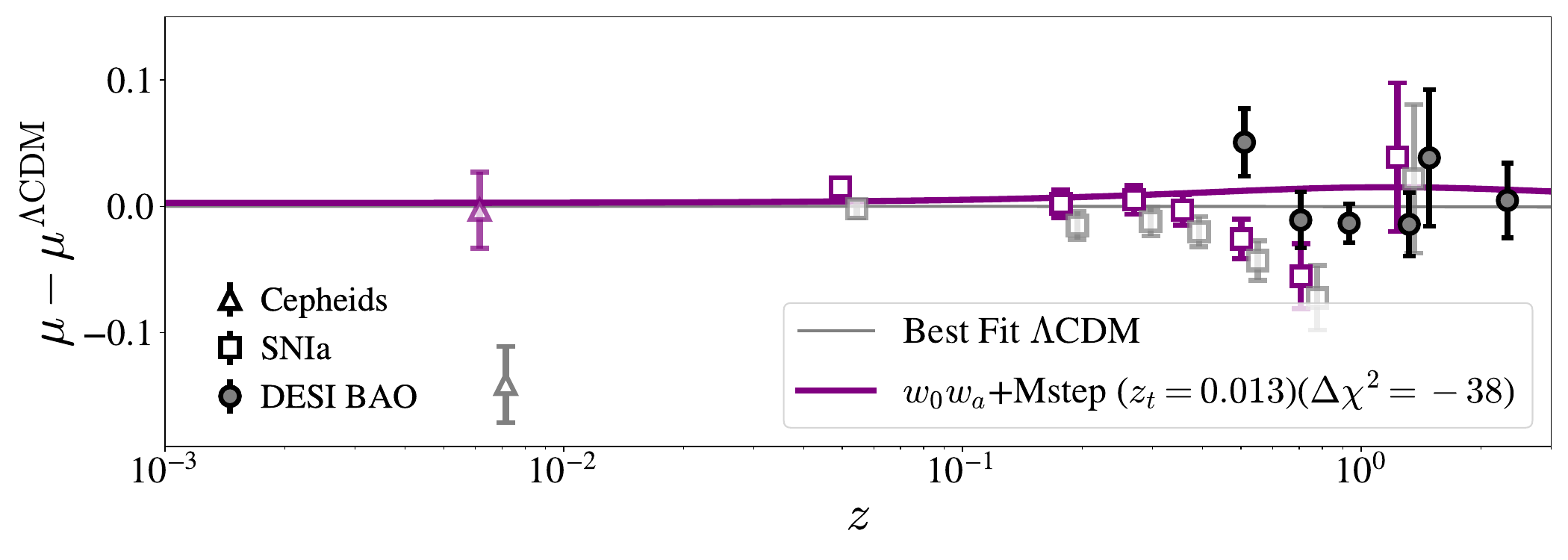}
    \caption{Similar as Fig.~\ref{fig:tension}, but for three phenomenological models, each compared to \lcdm. From top to bottom these are a model with a sharp step in the expansion rate (Hstep), one with a step in the absolute magnitude of SNIa (Mstep), and one with the \wowacdm\ expansion history {\it and} an Mstep. In each case, we provide in the legend the best-fit chi squared relative to \lcdm\ ($\Delta\chi^2$), as well as the best-fit redshift of the step ($\zt$). We also show data for the Cepheid calibrators, SNIa, and DESI BAO.  }
    \label{fig:Hstep}
\end{figure*}

\section{Results with specific models}\label{sec:results}

We next study specific cosmological models and their fit to data. 

Our principal results are shown in Table \ref{tab:results} and Fig.~\ref{fig:Hstep}. In  Table~\ref{tab:results}, we show the principal cosmological models (introduced in Sec.~\ref{sec:method}), and their respective goodness of fit to the SH0ES+SNIa+BAO+CMB data that was also introduced in that section.  Table~\ref{tab:results} is further divided into two sections, with the top part showing models with a very small improvement relative to \lcdm, while the lower section shows models with a substantial improvement of fit. In Fig.~\ref{fig:Hstep}, we show three select extended models and how they compare to \lcdm. In this Figure, we show the apparent magnitude relative to that in the \lcdm\ model, $\mu-\mu^{\Lambda \text{CDM}} $; in this case, SNIa data also shift between different models because of the change in the best-fit absolute magnitude; see Appendix~\ref{app:norm_details} for details. Throughout, we assume the combined {DESI DR2 + CMB + PantheonPlus+SH0ES} dataset, except in the second line of the top block of Table~\ref{tab:results}, where we show results for {PantheonPlus+SH0ES} alone to illustrate worsening of the fit when when {BAO + CMB} data are added to this combination.

Let us now interpret the $\Delta \chi^2$ values in Table~\ref{tab:results} and Fig.~\ref{fig:Hstep}. A striking feature of the \lcdm\ model, which is shown in all panels of Fig.~\ref{fig:Hstep} and listed in the top section of Table \ref{tab:results}, is that the Cepheid data point lies about $4$--$5\sigma$ away from the theoretical curves for both \lcdm\, and Hstep models. This arises because the absolute magnitude $M$ is treated as a free parameter, and its best-fit value is predominantly determined by high-redshift SNIa data points which adjust $M$ so as to reconcile their inferred distances with BAO measurements. As a result, the best-fit absolute magnitude is $M \simeq -19.4$, which is about $4$--$5\sigma$ away from the SH0ES calibrator value ($M = -19.243 \pm 0.027$). 

The top panel in Fig.~\ref{fig:Hstep} further shows that the situation is not alleviated in models that modify the low-redshift expansion history: for example, the Hstep model ($\chi^2=1498$, obtained for the illustrative transition redshift $\zt=0.013$) negligibly improves upon the fit in \lcdm\ model ($\chi^2=1500$). The same is true in the \wowa\ model ($\chi^2=1495$), which is not shown in Fig.~\ref{fig:Hstep} for clarity.  Interestingly, the model with spatial curvature improves the fit by $\Delta\chi^2=-10$ (so $\chi^2=1490$) for a single new degree of freedom ($\Ok$), but this improvement is much too small to explain the data when SH0ES are included in the dataset. All these (or other similar) low-z solutions do not improve the fit because a late-time transition in the Hubble parameter still leaves the SNIa as the dominant driver of $M$, pushing the best-fit magnitude to values that remain significantly offset from the Cepheid determination by SH0ES.

\begin{figure}[t]
    \centering
    \includegraphics[width=\linewidth]{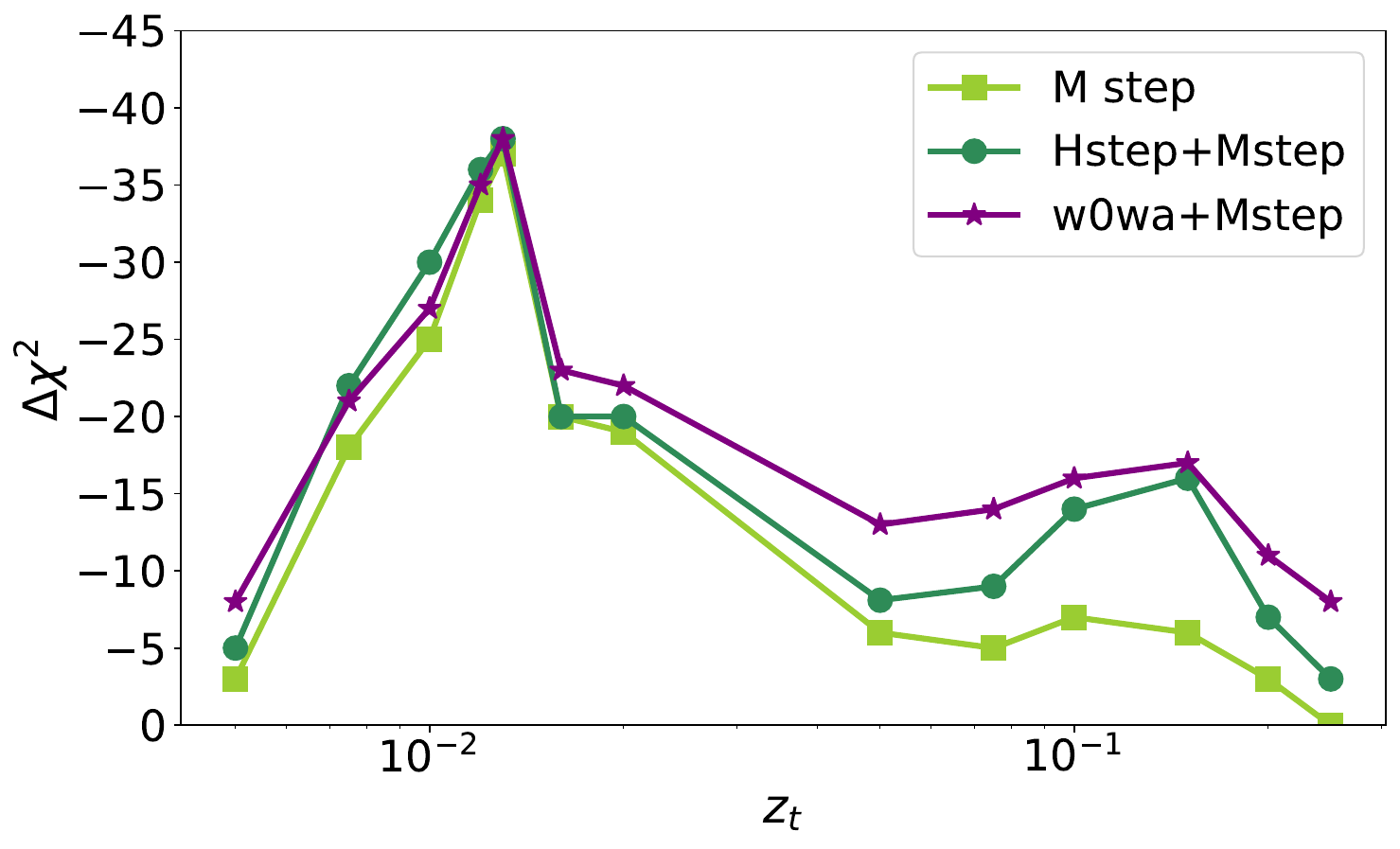}
    \caption{Improvement of the goodness of fit, $\Delta\chi^2$, relative to the best-fit \lcdm\ cosmological model, shown as a function of the transition redshift $\zt$. We show results for the Mstep model, Hstep+Mstep (assuming the same transition redshift in the Hubble parameter and absolute magnitude), and the Mstep model embedded in \wowa\ expansion history. In all three models, the largest improvement in the fit occurs with transition redshift of about 0.01. See text and Table \ref{tab:results} for more details.}
    \label{fig:chi2}
\end{figure}

In contrast, the lower section of Table \ref{tab:results} shows several models that have a much better fit than \lcdm\ (or Hstep) model. However, all of those models are trivial in that they effectively decouple the SH0ES $H_0$ measurement from the rest of SNIa Hubble diagram. First, we list the Mstep model, whose fit to data is better by a whopping $\Delta\chi^2\simeq -40$ relative to \lcdm. This model has two parameters relative to \lcdm, the pre-transition value of the absolute magnitude $\ME$ and the transition redshift $\zt$. All of the improvement in the fit of this model relative to \lcdm\ comes from improvement in modeling the PantheonPlus+SH0ES SNIa data, rather than BAO+CMB (to see this, compare the $\chi^2_{\mathrm{BAO+CMB}}$ value for this model to that of \lcdm; they are identical). Fig.~\ref{fig:chi2} shows that the best fit for the Mstep model is achieved at the transition redshift $\zt\simeq 0.013$. As expected, the best fit is therefore achieved at a very low transition redshift which effectively decouples Cepheids and SNIa in the SH0ES measurements. Table \ref{tab:results} also shows that $\ME$ in the Mstep model is very close to the best-fit $M$ value in the \lcdm\, model, allowing the SNIa distances to agree with BAO distances over the redshift range where they overlap. The late-time absolute-magnitude value, $M$, is close to the SH0ES-only  measurement (compare $M$ of the Mstep model to $M$ of the \lcdm\ model where PantheonPlus+SH0ES data alone were used -- the second model listed in Table~\ref{tab:results}). Thus the Mstep model avoids the massive penalties faced by the \lcdm\ and Hstep models. 

Next we show the Hstep+Mstep model; the two rows corresponding to this model in Table \ref{tab:results} assume respectively same or different redshift of transition in the absolute magnitude $M$ and Hubble parameter $H$ (that is, $\ztH$ and $\ztM$ respectively set to equal each other, or assumed as separate parameters). The two Hstep+Mstep model variants have a comparable fit, which is also similar to that of the Mstep model. Therefore, further introducing a step in $H$ does not help the fit once a step in $M$ has been adopted. The cyan curve in Fig.~\ref{fig:chi2} confirms that the behavior of the goodness of fit in the Hstep+Mstep model, as a function of the transition parameter (assuming the same transition redshift for $H$ and $M$), is very similar to that of the Hstep model, albeit with a slightly better goodness of fit, as would be expected for a model with an additional free parameter.

The last row of Table \ref{tab:results} shows the goodness of fit of the $w_0w_a$+Mstep model, which is the Mstep model that assumes the \wowa\ expansion history rather than \lcdm. We find that the additional freedom allowed by the \wowa\ expansion history does not appreciably change the fit of the Mstep model with \lcdm\ expansion history; the
former has $\Delta\chi^2=-43$ relative to the pure \lcdm\ model, while the latter has $\Delta\chi^2=-40$. Figure \ref{fig:chi2} confirms this, showing a slightly improved but overall similar fit as a function of the redshift at which $M$ transitions.


Fig.~\ref{fig:chi2} also identifies a second class of solutions that have a sharp step at $\zt\simeq 0.15$ though a much smaller peak in the likelihood than the $\zt\simeq  0.01$ solutions.
The former class of partially successful solutions is interesting because it is able to produce a higher $H_0$ with a relatively weaker $M$ transition even though the fit is modest. We discuss the reason why these $\zt\simeq 0.15$ solutions work in detail in Appendix \ref{app:weird_peak_z_015}.  

Overall then, we find that the principal feature required to fit data significantly better than \lcdm\ is to have a sharp, low-redshift transition in the absolute magnitude $M$. Such a transition, which is preferred by data at redshift $\zt\simeq 0.01$, effectively decouples the SH0ES calibration of the Hubble diagram, and allows the Hubble parameter to return to the value preferred by the CMB and BAO data, which is $H_0\simeq 67\kmsMpc$. Conversely, a sharp transition in the Hubble parameter, with or without additional freedom imposed by the \wowa\ model, does not appreciably improve the fit to the data relative to pure \lcdm.  

While effective in terms of fitting data, the Mstep model presents a rather trivial solution to the Hubble tension --- it effectively simply removes the SH0ES measurements from the data, which has been known for a long time to improve the overall fit to the combined data.

There is one more altogether different possibility for solving the Hubble tension, but much like the class of Mstep models it is trivial. This scenario is to break the Etherington duality relation that links the luminosity distance to angular diameter distance, $d_L(z)=(1+z)^2d_A(z)$. The Etherington relation is extremely robust in the standard cosmological model, and holds even the models with departures from the Friedmann-Lemaitre-Robertson-Walker (FRLW) metric \cite{Etherington:1933asu,Ellis:1973jva,Bassett:2003vu}. If this  relation were broken, however, than the BAO data would have an arbitrary normalization not tied to that from SNIa or the SH0ES measurement (so, the black points in Fig.~\ref{fig:tension} would be able to move vertically regardless of the red points). Etherington duality relation's breaking was recently investigated specifically in the context of Hubble tension by \citet{Teixeira:2025czm}. While the breaking of the duality relation remains a theoretical possibility to solve the Hubble tension, such a breaking would represent a radical (and, arguably, extremely unlikely) departure from the standard cosmological model.

\section{Inclusion of Surface Brightness Fluctuations Measurements}\label{sec:SBF}

In order to better  understand the dependence of our conclusions on the PantheonPlus+SH0ES dataset, we repeat the above analysis by including Surface Brightness Fluctuation (SBF) \cite{Tonry_Schneider}. SBF measure the pixel-to-pixel variance in a galaxy’s unresolved stellar light, which arises from Poisson fluctuations in the numbers and luminosities of stars per resolution element. The amplitude of these fluctuations decreases with distance and can therefore be used as a standard-candle–like distance indicator. Converting the observed fluctuation signal into a distance requires a calibration of the absolute SBF magnitude, which depends on the underlying stellar population and photometric zero point. 

We adopt the SBF measurements%
\footnote{\url{https://github.com/jjensen-uvu/sbf_distances_2021/tree/main}}
from Jensen et al.~\cite{Jensen_2025}. For this sample, the SBF measurements are calibrated based on James Webb Space Telescope (JWST) tip-of-the-red-giant-branch (TRGB) distances, themselves calibrated using the megamaser distance to NGC 4258, to determine a Cepheid-independent SBF zero point using Hubble Space Telescope observations. These SBF measurements alone give the constraint on the Hubble constant $H_0 = 73.8 \pm 0.7\,(\mathrm{stat}) \pm 2.3\,(\mathrm{sys})\,\kmsMpc$ \cite{Blakeslee_2021}.
Given the relatively large combined uncertainty, these measurements by themselves are not in significant tension with the combined BAO+CMB data. Nevertheless, it is still useful to repeat the above analysis including this dataset, as it provides low-redshift information on $H_0$ that is independent of the PantheonPlus+SH0ES sample.

Typically, these measurements are incorporated into the likelihood using the observed ``velocities'' and ``distances'' determined as described above. In this analysis, we use the group velocities listed in the \texttt{v\_grp} column, and the distance moduli and their associated uncertainties given in the \texttt{mM\_2025} and \texttt{err\_2025} columns, respectively, from the dataset presented in \cite{Jensen_2025}. To account for peculiar velocities, we assume an uncertainty of $\sigma_P = 250\,\kms$, as prescribed in \cite{Blakeslee_2021}, and add this contribution in quadrature to the uncertainties in the distance moduli after converting it to an equivalent uncertainty in distance modulus. The total uncertainty entering the likelihood for each data point is therefore given by
\begin{align}
    \sigma_{\mathrm{total}} = \sqrt{\sigma_\mu^2 + \left(\frac{5\,\sigma_P}{\ln 10\,\, \texttt{v\_grp}}\right)^2}\,.
\end{align}

To obtain a theoretical prediction for the distance, a simple approach is to use the Hubble law%
\footnote{The SBF dataset used here lies at very low redshift, $z \lesssim 0.02$, for which the relation $d=cz/H_0=v/H_0$ holds true.}
\begin{equation}
d_{\mathrm{theo}} = \frac{v_{\mathrm{obs}}}{H_0},
\end{equation}
and then convert $d_{\mathrm{theo}}$ into a distance modulus $\mu_{\mathrm{theo}}$, which can be directly compared to the observed distance moduli. In this approach, the only theory parameter entering the likelihood is $H_0$. However, this method is valid only for models that do not feature an $H$-transition. Since such models are considered in our analysis, we instead obtain a redshift from the group velocity by solving the following cosmographic cubic equation 
\begin{equation}
\texttt{v\_grp}=cz\left(1-\frac{z}{2}(1+q_0)+\frac{z^2}{6}(2+4q_0+3q_0^2-j_0)\right),
\end{equation}
and only consider the positive roots which are less than 1 in value. In the above equation we choose the value of the acceleration parameter, $q_0=1/2 (\Omega_m-2\Omega_\Lambda)=-0.595$, and the jerk parameter, $j_0=1$, corresponding to the value of $\Omega_m=0.27,\,\Omega_\Lambda=0.73$ as used in the original COSMICFLOWS-3 \cite{Tully_2016} database, from which these group velocities were originally derived.
We then compute the comoving distance $d_M(z)$ and subsequently the luminosity distance $d_L(z)=(1+z)d_M(z)$ for a given theoretical model and convert it into the distance modulus.

To account for systematic uncertainties in the zero-point calibration, we introduce a free parameter $\mathcal{Z}_\mathrm{SBF}$ with a Gaussian prior $\mathcal{N}(0, 0.063^2)$, which is added directly to the theoretical prediction of the distance modulus. The width of this prior is chosen to match the total systematic uncertainty in the TRGB calibration of Hubble Space Telescope SBF measurements, as described in \cite{Jensen_2025}.

Having described the construction of the SBF likelihood, we now present the results obtained after including these measurements in our analysis. We consider two scenarios: (A) fixing $\mathcal{Z}_\mathrm{SBF} = 0$, and (B) allowing $\mathcal{Z}_\mathrm{SBF}$ to vary with the prior given above. The first case corresponds to a situation with no systematic uncertainty in the SBF zero-point, and therefore in $H_0$. Table~\ref{tab:results_sbf} summarizes the results for both scenarios.

\begin{table*}[t]
\centering
\renewcommand{\arraystretch}{1.2}
\setlength{\tabcolsep}{10pt}
\begin{tabular}{l P{1.4cm} c c c c c}
\midrule
\multicolumn{7}{c}{\textbf{Baseline+SBF} (zero-point uncertainty ignored)} \\
\midrule
\textbf{Model} & $\boldsymbol{\chi^2_{\mathrm{total}}}$ &  \cellcolor{deltared}
$\boldsymbol{\Delta\chi^2}$ &$\boldsymbol{\chi^2_{\mathrm{BAO+CMB}}}$ & $\boldsymbol{\chi^2_{\mathrm{Panth+SH0ES}}}$ & $\boldsymbol{\chi^2_{\mathrm{SBF}}}$ & $\Delta$DOF \\
\midrule
\addlinespace
\addlinespace
\textbf{LCDM} & 1645 & \cellcolor{deltared}0 & 32 & 1478 & 135 &0  \\
\addlinespace
\textbf{LCDM} (SBF only) & -- &\cellcolor{deltared}-- & -- & -- & 97 &0 \\
\addlinespace
\textbf{HStep} & 1630 & \cellcolor{deltared}$-15$ &24 & 1498 &108 &2 \\
\addlinespace
\textbf{MStep} & 1614 &\cellcolor{deltared} $-31$ &26 & 1446 & 142 &2 \\
\addlinespace
\textbf{HStep+MStep} ($\ztH=\ztM$) & 1596 &\cellcolor{deltared} $-49$ &19 & 1465 &112 &3 \\
\midrule
\addlinespace
\multicolumn{7}{c}{\textbf{Baseline+SBF+$\mathcal{Z}_\mathrm{SBF}$} (zero-point uncertainty included)} \\
\midrule
\textbf{Model} & $\boldsymbol{\chi^2_{\mathrm{total}}}$  & \cellcolor{deltablue}$\boldsymbol{\Delta\chi^2}$ & $\boldsymbol{\chi^2_{\mathrm{BAO+CMB}}}$ & $\boldsymbol{\chi^2_{\mathrm{Panth+SH0ES}}}$ & $\boldsymbol{\chi^2_{\mathrm{SBF}}}$ & $\Delta$DOF \\
\midrule
\addlinespace
\addlinespace
\textbf{LCDM} & 1574 &\cellcolor{deltablue}0 & 18 & 1483 & 73 &0   \\
\addlinespace
\textbf{LCDM} (SBF only) & -- &\cellcolor{deltablue} -- & -- & -- & 71 &0 \\
\addlinespace
\textbf{HStep} & 1571 & \cellcolor{deltablue} $-3$ & 18 & 1482 & 71 &2 \\
\addlinespace
\textbf{MStep} & 1533 &\cellcolor{deltablue}   $-41$ & 17 & 1444 & 72 &2 \\
\addlinespace
\textbf{HStep+MStep} ($\ztH=\ztM$) & 1530   & \cellcolor{deltablue} $-44$ & 17 & 1442 & 71 &3 \\
\bottomrule
\end{tabular}
\caption{Goodness of fit for our models, shown for the scenario when surface brightness fluctuations data are added to the fiducial datasets (studied and displayed in  Table \ref{tab:results}). The top half of the table shows the assumption of a fixed zero-point uncertainty of the SBF, while the bottom half shows a more realistic scenario when this uncertainty has been varied in the analysis.  The total $\chi^2$ is shown for the baseline data consisting of DR2 BAO+CMB+PantheonPlus+SH0ES+SBF, along with its contributions from BAO+CMB data, PantheonPlus+SH0ES data, and SBF data alone. The column labeled $\Delta\chi^2$ shows the difference of the goodness of fit relative to the fiducial \lcdm\ model. To enable a direct, idealized comparison of the impact of including SBF data, the relevant quantities are the blue $\Delta\chi^2$ values in this table compared to those in Table~\ref{tab:results} for the corresponding models.  The last column displays the number of degrees of freedom of a given model relative to \lcdm. }
\label{tab:results_sbf}
\end{table*}

We first discuss the results corresponding to SBF measurements without a zero-point uncertainty, shown in the upper half of Table~\ref{tab:results_sbf}. As discussed earlier, in this case the SBF data are in tension with BAO+CMB at a level similar to that of the PantheonPlus+SH0ES data. Therefore, when SBF distance measurements are combined with BAO+CMB in the $\Lambda$CDM model, the fit to the SBF data is expected to worsen, which is indeed seen in the first two rows of the table. We next consider the Hstep model and find a moderate improvement in the overall fit, with $\Delta\chi^2 = -15$ relative to $\Lambda$CDM. From the individual $\chi^2$ contributions, it is clear that most of this improvement is driven by the SBF data. This is expected, since at the likelihood level SBF distances depend only on the value of $H_0$. The Hstep model can therefore simultaneously fit the BAO+CMB data with $H_0 \sim 68$ \textit{and} the low-redshift SBF data with $H_0 \sim 73$. We then move on to the Mstep model. Although this model leads to a larger overall improvement, with $\Delta\chi^2 = -31$ compared to $\Lambda$CDM, it does not improve the fit to the SBF data at all. This is again expected, as SBF distances are insensitive to an absolute-magnitude transition in their likelihood (as they measure distances directly). Finally, we consider the Hstep+Mstep model. This model provides the best overall fit, with $\Delta\chi^2 = -49$ relative to $\Lambda$CDM. Since Hstep+Mstep allows for both an $H_0$ transition which is preferred by the SBF data (when combined with the BAO+CMB data), and a magnitude transition which is preferred by the Cepheid data, it leads to an improved fit to both the SBF and PantheonPlus+SH0ES datasets compared to the $\Lambda$CDM model.

We now turn to the second case, in which a variable zero-point uncertainty in the SBF measurements is included. These results correspond to the lower half of Table~\ref{tab:results_sbf}. Now, we find that the level of improvement obtained using the various models does not change essentially at all compared to the case without SBF, in complete distinction from the case with a fixed zero-point uncertainty. This is further reflected in the values of $\chi^2_{\mathrm{SBF}}$, which remain nearly unchanged across the different models we consider. 

This distinct insensitivity of the SBF likelihood with a varying zero-point uncertainty to the models that we explored can be understood as follows. Allowing for a free zero-point uncertainty effectively increases the error bars on the inferred distances. As a result, any changes introduced by non-$\Lambda$CDM models can be accommodated within these enlarged uncertainties. Consequently, once the zero-point uncertainty is included, the SBF data yield very similar constraints across the different models.

Summarizing this section, we conclude that, in the realistic case when the zero point uncertainty of the SBF has been varied, the addition of SBF to our fiducial dataset consisting of baryon acoustic oscillations, type Ia supernovae, and SH0ES measurements, is minimal. This analysis also highlights the potential of SBF distances to provide an independent handle on the Hubble tension. As the calibration of SBF distances improves through an expanded sample of TRGB–SBF calibrators, the associated zeropoint systematic uncertainty is expected to decrease. This combination of more precise SBF distances combined with the PantheonPlus+SH0ES SNIa dataset will help test and potentially rule out these admittedly artificial Hstep and Mstep solutions which appear to work for the Hubble tension.

\begin{figure*}[t]
    \centering
    \includegraphics[width=\linewidth]{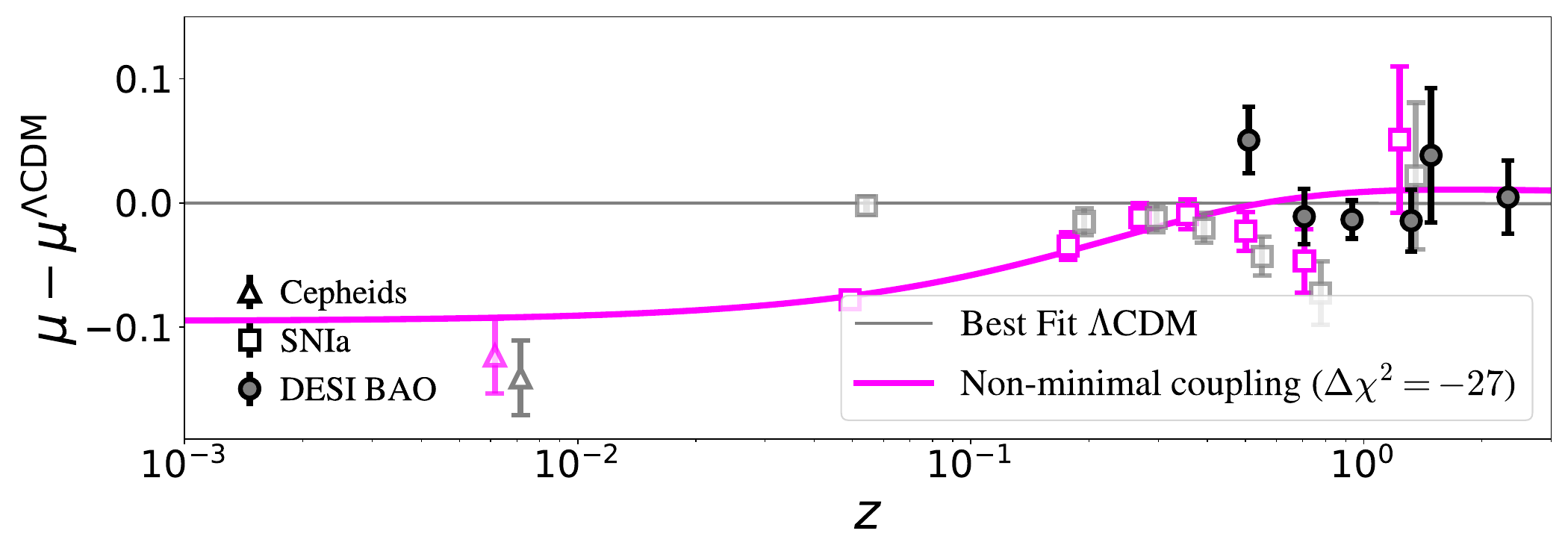}
    \caption{Similar as Figs.~\ref{fig:tension} and \ref{fig:Hstep}, but for the physical scalar-field model discussed in Sec.~\ref{sec:Wolf_model}. The fit of the model is acceptable ($\Delta\chi^2\simeq -27$ relative to \lcdm)  due to its ability to change the absolute magnitude of SNIa. While not as good a fit to the data as models with a very low-redshift ($\zt\simeq 0.01$) transition, the model's reasonable fit is due in part to the fact that the transition at $\zt\sim 0.15$ is gradual, as it fits the data  in the intermediate redshift range better than models with a sharp transition. 
    See text for more details.}
    \label{fig:wjw}
\end{figure*}

\section{Worked Physical Example: Non-minimally coupled scalar field}\label{sec:Wolf_model}

We now provide a specific physical example of a model that features a transition in the Hubble parameter. We do so to balance the previous discussion which was entirely based on models that are purely phenomenological, and extreme in the sense of the sharpness of transitions in the relevant parameters (Hubble parameter and the absolute mangitude of SNIa). 

We consider the non-minimally coupled scalar field model introduced in \citet{Wolf:2025jed,Wolf_2025}. This model is particularly appealing because it possesses sufficient flexibility to reproduce the expansion history favored by the combination of DESI DR2 BAO, CMB, and uncalibrated SNIa data. Moreover, it inherently features a transition in the effective gravitational constant due to the coupling between the scalar field and gravity. This gravitational transition provides a physical motivation for considering an analogous transition in the SNIa absolute magnitude, as discussed in the following sections.  

We begin by outlining the action for this model:
\begin{equation}
\begin{aligned}
S = \int d^4x \, \sqrt{-g} \Bigg[
    &\frac{M_{\text{Pl}}^2}{2} F(\varphi) R
    - \frac{1}{2} G(\varphi) X
    - V(\varphi) \\[4pt]
    &- J(\varphi) X^2
    + \mathcal{L}_M (g_{\alpha\beta}, \psi_M)
\Bigg],
\end{aligned}
\end{equation}
where \( g_{\alpha\beta} \) is the metric, \( R \) the Ricci scalar, \( \varphi \) the scalar field, 
\( X = \partial_\mu \varphi \partial^\mu \varphi \), and \( \mathcal{L}_M \) denotes the matter Lagrangian. For the non-minimally coupled scalar field, the relevant functions are expanded as
\begin{equation}
\begin{aligned}
F(\varphi) &\simeq 1 - \xi \frac{\varphi^2}{M_{\text{Pl}}^2}, \\[6pt]
V(\varphi) &\simeq V_0 + \beta \varphi + \frac{1}{2} m^2 \varphi^2,
\end{aligned}
\end{equation}
with \( G(\varphi) = 1 \) and \( J(\varphi) = 0 \).  

The non-minimal nature of this model arises through the \( F(\varphi) \) term, which modifies the coupling between the scalar field and the Ricci scalar \( R \). In contrast to General Relativity (GR) or the minimally coupled models, where this coupling is unity, the scalar field dependence of \( F(\varphi) \) introduces a dynamical effective Planck mass, leading to a time-dependent gravitational constant. The model introduces four additional parameters relative to \(\Lambda\)CDM: \( \xi, V_0, \beta,\) and \( m \) (the field mass). However, only three are independent since \( V_0 \) is adjusted to ensure that the sum of all fractional energy densities equals unity.  

This coupling implies that the effective gravitational constant varies over cosmic time, \( G_\mathrm{eff}(t) \), driven by the evolution of the scalar field. Such variation motivates the consideration of a corresponding time-varying absolute magnitude for SNIa. Although several examples of ``\( G \)-induced'' transitions in SNIa luminosities have appeared in the literature \cite{Marra:2021fvf, Gupta:2021tma, Ruchika:2023ugh, Amendola:1999vu,Garcia-Berro:1999cwy,PhysRevD.97.083505}, the underlying physical mechanisms remain a subject of ongoing debate.  

To remain agnostic regarding specific implementations, we do not adopt any single model from the literature. Instead, we propose a generalized phenomenological link between the gravitational transition in the non-minimally coupled scalar field model and a corresponding SNIa absolute magnitude transition. This connection is implemented through the modified gravity function \( \mu_\mathrm{MG}(z) \) \citep{Ishak:2024jhs}, which parametrizes deviations from the Poisson equation as a time-dependent rescaling of Newton’s constant. Neglecting any scale dependence of \( \mu_\mathrm{MG} \), and assuming the SNIa luminosity scales as a power law with the gravitational constant \( G \), the absolute magnitude can be expressed as a logarithmic function of \( \mu_\mathrm{MG}(z) \):
\begin{equation}
M(z) = \ME + n \log[\mu_\mathrm{MG}(z)],
\end{equation}
where the parameter \( n \) encodes the power-law index, and \( \ME \) denotes the asymptotic SNIa magnitude at high redshift (\( z > 1 \)).



We implemented this model by modifying \texttt{hi\_class}\cite{Zumalacárregui_2017} and used \texttt{cobaya}\cite{Torrado:2020dgo} to run the minimiser \texttt{iminuit}\cite{James:1975dr} using the combined DESI DR2 BAO + CMB + PantheonPlus + SH0ES data.
We find that the best-fit to this model with a non-minimally coupled scalar field and a time-varying $G$  gives a $\dchisq=-27$ over the best-fit \lcdm. This is further illustrated in Fig.~\ref{fig:wjw} which shows the Hubble diagram corresponding to the best-fit non minimal coupling model along with the data used. The success of the model comes from its ability to accommodate the change in the absolute magnitude $M$, which has been implemented via a time-varying gravitational constant. [It is useful to again refer to Fig.~\ref{fig:tension} to understand what is needed for a successful model.] Specifically, we find that the best-fit model favors a transition relative to \lcdm\ of $\Delta M\simeq -0.1$ that smoothly takes place over a range of redshifts around $\zt\simeq 0.1$--$0.2$. Note that the non-minimally coupled scalar field model \textit{without} the $M$ transition would only achieve $\dchisq\simeq -10$ relative to \lcdm, again illustrating the importance of the absolute-magnitude transition. Conversely, the fact that the goodness of fit of the model with a smooth $M$ transition is $\dchisq=-27$ rather than $\simeq -40$ as with the Mstep (or Hstep+Mstep) models discussed in Sec.~\ref{sec:results} is that the scalar-field model's transition is gradual and at an intermediate redshift ($\zt\sim 0.15$), rather than sharp and at a much lower redshift ($\zt\simeq 0.01$), the latter of which would better fit data that includes SH0ES. This is further discussed in Appendix \ref{app:weird_peak_z_015}.

We have identified one more interesting feature of the scalar-field model studied in this section: its best-fit value of the Hubble constant is $\simeq 72\kmsMpc$, even higher than that in models with a sharp step in $M$ with transition specified at $\zt\simeq 0.15$ (which give $H_0\simeq 70\kmsMpc$; see Table \ref{tab:results2}). The reason for this high $H_0$ is the fact that the gradual transition in the expansion history enabled by the scalar-field model fits data better than the sharp transition at $\zt\simeq 0.15$ in our Hstep+Mstep model; this can be seen by comparing Fig.~\ref{fig:wjw} with the lower panel of Fig.~\ref{fig:zt_0.01_or_0.15}. The scalar-field model's gradual $H$ transition also fits data better than the $\zt\simeq 0.15$ Mstep model even when the latter is embedded in a \wowacdm\ background. This is further discussed in Appendix \ref{app:weird_peak_z_015}, while Table  \ref{tab:results2} has more details about the derived parameters of the scalar-field model compared to two phenomenological step-models, as well as \lcdm.

Overall, the scalar-field model with a non-minimal coupling and an $M$ transition  serves a useful purpose of illustrating what precisely is required in order to fit the combination of SH0ES and other cosmological data. It could be that a model like this, perhaps better motivated by compelling theory or else favored by other data in cosmology, will prove to be useful in explaining the Hubble tension.

\section{Conclusions} \label{sec:concl}

We have investigated the notion that low-redshift modifications to expansion history face very steep challenges to resolve the Hubble tension. To better understand this thesis, we studied several simple phenomenological models that modify expansion history and/or introduce a jump in the absolute magnitude of Type Ia supernovae. In order to gain more understanding about what the data are indicating, we also studied a physical model of dark energy that is fine-tuned for the purposes of explaining the Hubble tension. We studied how these models fit the combined SH0ES+SNIa+BAO+CMB data. Instead of adopting the $M$-prior or $H_0$-prior  commonly found in the literature, we instead employed the full SH0ES data in order to properly quantify the corresponding $\Delta\chi^2$ values. We also clarified some subtleties associated with using an $M$ prior, and emphasized that care must be taken to avoid double-counting information when such a prior is applied.

We found that a sharp transition in the expansion rate does not appreciably improve the fit to data relative to the standard \lcdm\ model; this remains true even when allowing a more flexible expansion history accommodated by the \wowa\ model. This occurs because, when the model is fit to the BAO data calibrated by Planck \lcdm\, sound horizon, the resulting expansion anchors the normalization of the SNIa in order to agree with the BAO and the theoretical model, and this normalization anchors the absolute magnitude $M$ to a value that is approximately $5\sigma$ away from that preferred by the Cepheid calibrators. 

In contrast, a sharp step in the absolute magnitude of Type Ia supernovae at redshift $\zt\simeq 0.01$ works much better; it improves the overall fit to the  data by $\Delta\chi^2\simeq -40$. It is easy to derive a back-of-envelope value of the magnitude step preferred by these models: if we write the SNIa apparent magnitude as $m = 5\log_{\rm 10}(H_0d_L(z)) + \mathcal{M}$, where $d_L$ is the luminosity distance at a given redshift, then the nuisance parameter that combines the absolute magnitude $M$ and the Hubble constant $H_0$ is defined by \citep[e.g.][]{Huterer:2023mmv}
\begin{equation}
    \mathcal{M}\equiv M - 5\log_{\rm 10}(H_0\times 1\Mpc)+25.
\end{equation}
Then the desire to keep $\mathcal{M}$ to whatever SNIa measurements favor it to be, but to somehow lower $H_0$ from $\sim 67\kmsMpc$ that is favored at higher redshifts to $\sim 73\kmsMpc$ favored at $z\rightarrow 0$, immediately implies that $M$ should increase by about $0.15$--$0.18$ magnitudes. This is indeed the jump that the various Mstep models favor (see the typical differences between $M$ and $\ME$ in Table \ref{tab:results}).

However these "Mstep" solutions are trivial, as they effectively decouple the SH0ES measurements and allows an arbitrary normalization of SNIa Hubble diagram. This is similar to another known possibility for explaining the Hubble tension, the decoupling of the Etherington distance duality relation $d_L=(1+z)^2d_A(z)$; it too is trivial as it allows for a separate anchoring of SNIa measurements relative to those from the CMB and BAO. 

We also found that  models with a transition in the absolute magnitude $M$ at an intermediate redshift, $\zt\simeq 0.15$, while not as successful as models with the transition at very low redshift, nevertheless  lead to a somewhat acceptable fit to the data ($\dchisq\simeq -16$ relative to \lcdm). The reason that the physical scalar-field model, which also accommodates the $M$ transition at intermediate redshift, does better ($\dchisq\simeq -27$) is because its $M$ and Hubble transitions are more gradual and the data mildly prefer that. Interestingly, scalar-field models with an $M$ transition also result in the Hubble constant of $H_0\simeq 72\kmsMpc$, which is among the highest among the models that we studied.

We repeated the above analysis by including the surface brightness fluctuations (SBF) dataset, which is calibrated independently from the SH0ES Cepheid distance scale. Given the current level of uncertainty in these measurements, the inclusion of SBF distances does not lead to a significant change in our principal conclusions. In an idealized scenario where systematic uncertainties are neglected, we find that improved SBF precision can have a noticeable impact on the results and may help further test magnitude-transition–type solutions to the Hubble tension.

Finally, we have also studied a previously proposed physical model of dark energy with the non-minimal coupling to gravity. This model by itself could not resolve the Hubble tension and fit the combined data because, as discussed above, even an arbitrary change to the low-z expansion history is insufficient to succeed in this. However, when ``fortified" by the late-time temporal change in Newton's constant that is conjectured to lead to a jump in the Chandrasekhar mass and hence the absolute magnitude of SNIa, the model succeeds in providing a reasonably good fit to the data (the aforementioned $\dchisq\simeq -27$ relative to \lcdm). 

Despite such partial successes, our general conclusion is that the only models that fit data much better than \lcdm\ are the ones with an $M$ transition at very low redshift,  $\zt\simeq 0.01$. Those models are trivial in that they explicitly decouple of the SH0ES measurements of the anchors of the distance ladder from the higher-redshift SNIa and BAO measurements. A similar conclusion applies to models that induce breaking of the Etherington relation which links the angular-diameter distance and luminosity distance; this relation is considered extremely robust and holds even in beyond-FLRW models, and its breaking would also resolve the Hubble tension trivially as it would decouple the SNIa Hubble diagram from that of the BAO. An even more far-fetched low-redshift solution to the Hubble tension would be to postulate a "just-so" expansion history where the theoretical prediction for the distance-redshift relation rapidly wiggles to pass through the SNIa and BAO data in the overlapping redshift range where they have different overall normalizations (see the shaded region in Fig.~\ref{fig:tension}). These conclusions are unlikely to be changed in models that allow an even more flexible smooth late-universe expansion history, as those models have fits that are comparable to, and not much better than, \wowa\ (e.g.\ \cite{Bansal:2025ipo}).

\vspace{0.5cm}
\textit{Acknowledgments.}
This work has been supported by the Department of Energy under contract DE‐SC0019193. PB thanks William Wolf for several useful discussions related to the non-minimal coupling model and its implementation in \texttt{hi\_class}. We also thank John Blakeslee, Pedro Ferreira, Vivian Poulin, Adam Riess, and Dan Scolnic for useful discussions and comments on earlier versions of the manuscript.

\appendix

\section{On the subtleties of using $M$ prior}
\label{app:Mprior}

In the analysis above, we have used the full PantheonPlus+SH0ES dataset, which consistently accounts for the Cepheid-calibrated supernovae. In the literature, however, it is common to adopt a Gaussian prior on the absolute magnitude of SNe~Ia, $M$.

Care must be taken when employing this prior in order to avoid double-counting. The widely used prior, $M = -19.253 \pm 0.027$, comes from the SH0ES collaboration. It is important to recall how this value is derived. In their analysis to constrain $H_0$, the SH0ES team does not use the full high-redshift PantheonPlus sample, but instead restricts to SNe~Ia at $z < 0.15$, where the statistical weight is sufficient. From this low-redshift sample, they obtain $H_0 = 73.04 \pm 1.04$ and $M = -19.253 \pm 0.027$ \cite{Riess:2021jrx}.  

\begin{figure}[t]
    \centering
    \includegraphics[width=\linewidth]{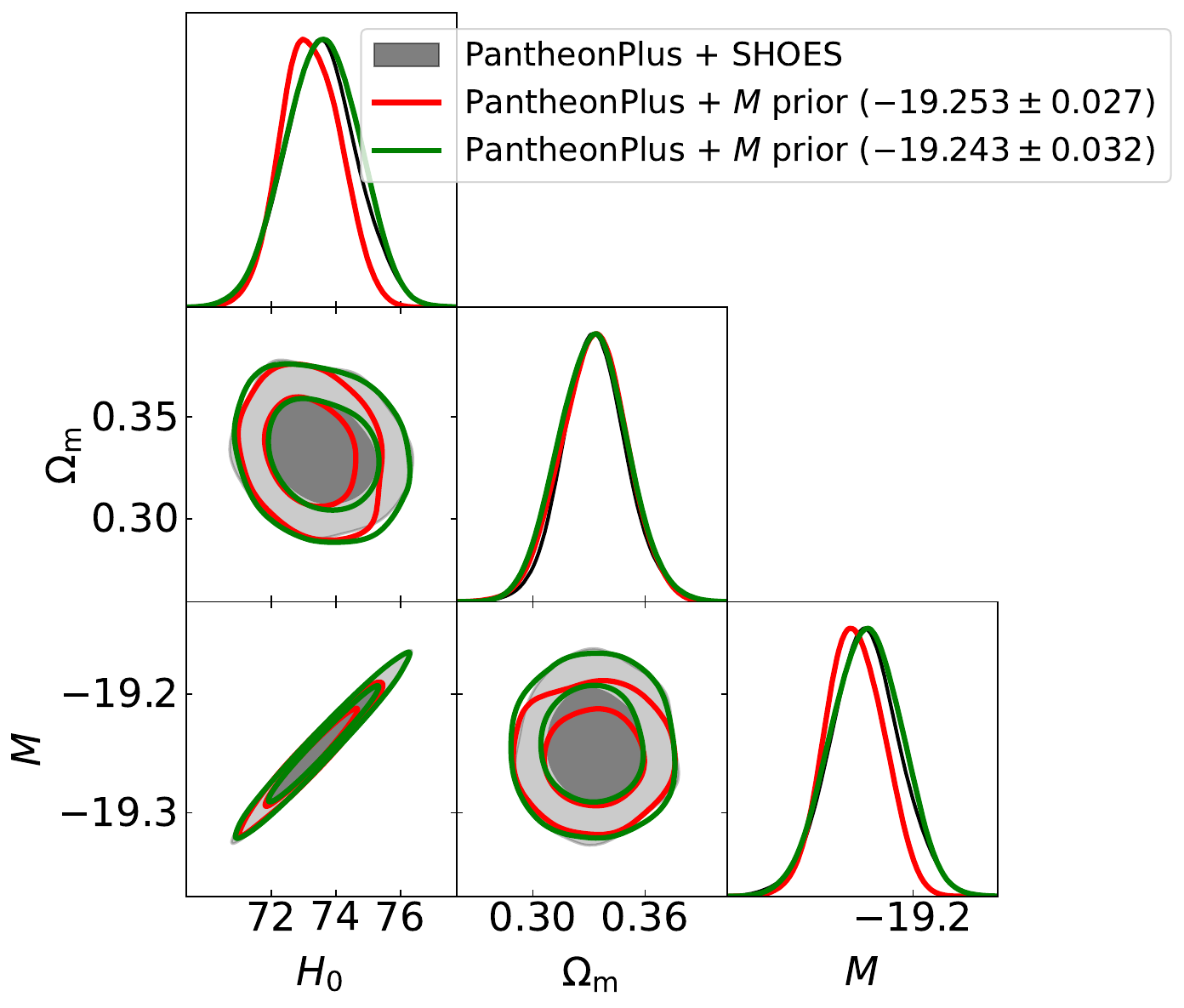}
    \caption{Tests of the replacement of the SH0ES dataset with an equivalent $M$-prior. We show constraints on the Hubble constant, matter density relative to critical,  and absolute maggnitude of SNIa in the \lcdm\ model. The solid grey contours show the fiducial case, the red contours show the case when the $M$ information is effectively double-counted, while the green contours show the corrected $M$ prior that approximately matches the fiducial constraints. See text for details.}
    \label{fig:mprior}
\end{figure}

When this $M$ prior is applied together with the full PantheonPlus sample (often with a low-redshift cut of $z > 0.01$), it effectively double-counts the SNe in the range $0.01 < z < 0.15$. This double counting manifests in Fig.~\ref{fig:mprior}, where constraints obtained from PantheonPlus plus the $M$ prior (red contours) are artificially tighter compared to those from the full PantheonPlus+SH0ES dataset. To avoid this issue, the error on the SH0ES-derived magnitude can be inflated. This was done by \citet{Mprior} using a procedure called ``de-marginalization,'' yielding an adjusted prior of 
\begin{equation}
    M = -19.243 \pm 0.032.
\end{equation}
As shown in Fig.~\ref{fig:mprior}, adopting this inflated-error prior (green contours) leads to constraints consistent with those from the full PantheonPlus+SH0ES analysis.  

We again emphasize that we did not need to impose any prior on $M$ to represent SH0ES data, as we used the full PantheonPlus+SH0ES dataset; the discussion above is therefore mostly pedagogical. 

\begin{table*}[t]
\centering
\renewcommand{\arraystretch}{1.2}
\setlength{\tabcolsep}{10pt}
\begin{tabular}{l P{1.4cm} c c c c c c}
\midrule
\multicolumn{7}{c}{\textbf{Modest fit to DR2 BAO + CMB + PantheonPlus + SH0ES data}} \\
\midrule
\textbf{Model} & $\boldsymbol{\chi^2_{\mathrm{total}}}$ &  \cellcolor{deltablue}$\boldsymbol{\Delta\chi^2}$ & $H_0$ ($\HnodE$) &$\Omega_m$ & $\ME$ & $M$ & $\Delta$DOF \\
\midrule
\addlinespace
\textbf{\lcdm}  & 1500 &\cellcolor{deltablue} $0$ &68.71 & 0.297 & --- & $-19.399$ & 0 \\
\addlinespace
\textbf{Hstep+Mstep}  ($\ztH=\ztM=0.15$) & 1484 & \cellcolor{deltablue} $-16$ & 70.24 (67.94) & 0.305 & $-19.391$ & $-19.341$ & 3 \\
\addlinespace
\textbf{$w_0w_a$+Mstep} ($\zt=0.15$)& 1484 &  \cellcolor{deltablue}$-16$ & 69.99 & 0.291 & $-19.383$ & $-19.352$ & 4 \\
\addlinespace
\textbf{Non-minimal Coupling} & 1473 & \cellcolor{deltablue}  $-27$ & 71.78 & 0.272 & $-19.42$ & $-19.281$ & 4 \\
\bottomrule
\end{tabular}
\caption{Goodness-of-fit values and associated parameters for \lcdm\, two purely phenomenological modes (Hstep+Mstep and \wowa+Mstep), as well as the scalar field model with non-minimal coupling. The cosmological parameters shown are the Hubble constant, matter density relative to critical, pre-transition and post-transition SNIa absolute magnitudes, and degrees of freedom relative to \lcdm. In the Hstep+Mstep model, we additionally show the pre-transition Hubble-constant value, $\HnodE$. A noteworthy feature is the Hubble constant value in these models; see text for more details.}
\label{tab:results2}
\end{table*}

\section{More details about the normalizations in Figs.~\ref{fig:tension} and \ref{fig:Hstep}}
\label{app:norm_details}

At first glance it may seem surprising that the SNIa data points in Figs.~\ref{fig:tension} and \ref{fig:Hstep} shift depending on the theoretical model. This arises from how the observational quantities are represented, so we outline here precisely what is plotted. For the SNIa data points, we display  
\begin{equation}
\mu - \mu^{\Lambda \text{CDM}} = m^{\text{data}} - M^{\text{th.,best-fit}} - \mu^{\Lambda \text{CDM}},
\end{equation}
where $\mu$ is the distance modulus (and $\mu=m-M$ in general) and $M^{\text{th.,best-fit}}$ is the absolute magnitude that is determined by a fit of the corresponding theoretical model considered in the figures to SNIa data. This makes the SNIa points explicitly model-dependent, since the absolute magnitude $M^{\text{th.,best-fit}}$ is itself theory-dependent.  

The situation is different for the Cepheid calibrator data point shown in these Figures. Because the Cepheids only measure the absolute magnitude (which we will call $M^{\text{data}}$), their corresponding distance modulus -- so $\mu$ in the expression for the oordinate $\mu-\mu^{\Lambda \text{CDM}}$ -- will necessarily be theory-dependent. Despite this unusual feature of the 
measurement varying with the theory model, such a procedure will allow for a fair comparison between Cepheid measurements and individual theory models.  For the Cepheid calibrators we therefore plot the theory distance modulus, but one where the theory/best-fit absolute magnitude $M^{\text{th.,best-fit}}$ has been replaced by the measured value $M^{\text{data}}$
\begin{align}
& (\mu^{\text{theory}} +M^{\text{th.,best-fit}} - M^{\text{data}}) - \mu^{\Lambda \text{CDM}}. 
\label{eq:mu_cepheids}
\end{align}
An immediate consequence, evident from Eq.~(\ref{eq:mu_cepheids}), is that if $M^{\text{theory}} = M^{\text{data}}$, then the calibrator data points reduce to $\mu^{\text{theory}} - \mu^{\Lambda \text{CDM}}$, exactly matching the theoretical prediction. Conversely, the larger the difference between $M^{\text{theory}}$ and $M^{\text{data}}$, the greater the offset of the Cepheid data points from the theory curve. This behavior is consistent with how Cepheids are treated within the PantheonPlus+SH0ES likelihood.

For the BAO data points, we proceed as follows: we first compute the sound horizon using the best-fit $\Lambda$CDM cosmology, then use this to obtain the transverse comoving distance from Table~IV of Ref.~\cite{DESI:2025zgx}. These comoving distances are then converted to luminosity distances, and hence to distance moduli, assuming the validity of the distance--duality relation.  

\section{Partial success of models with step at $z\simeq 0.15$}
\label{app:weird_peak_z_015}

\begin{figure*}[t]
\includegraphics[width=0.9\linewidth]{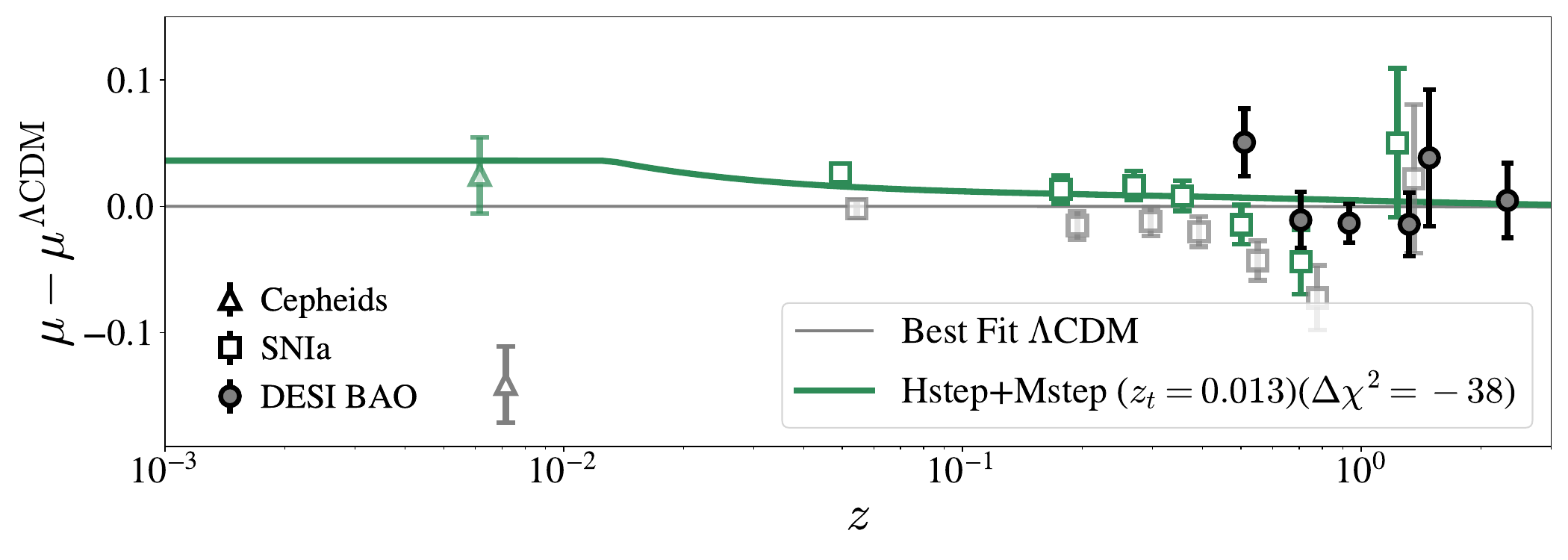}
\includegraphics[width=0.9\linewidth]{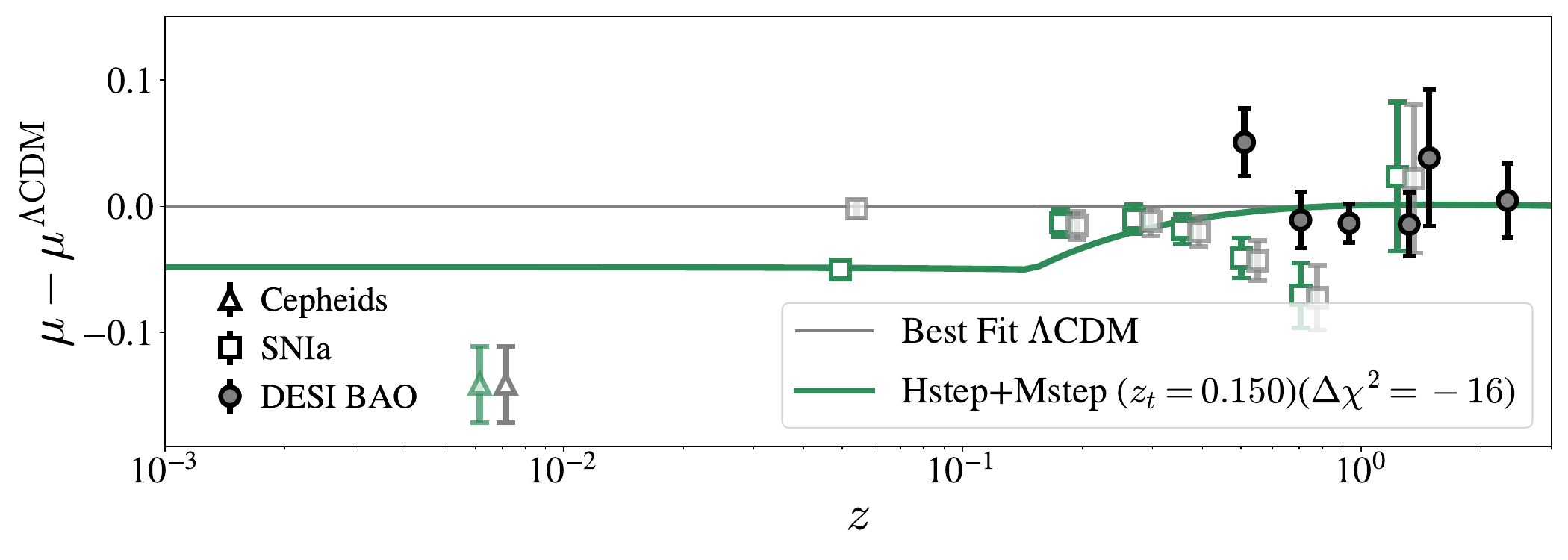}
    \caption{Similar as Figure \ref{fig:Hstep}, except providing a direct comparison of a single model with a sharp $H$ and $M$ transition at two alternative redshifts: $\zt=0.013$ (top panel), and $\zt=0.15$ (bottom panel). The former case provides a much better fit to the data, as indicated in respective legends.   }
    \label{fig:zt_0.01_or_0.15}
\end{figure*}

We now comment on some interesting features of models that feature an absolute magnitude transition \textit{not} at a very low redshift, but rather somewhat earlier in the expansion history, at $\zt\simeq 0.15$. 

As discussed in Sec.~\ref{sec:results}, an $M$ transition at very low redshift ($z \sim 0.01$) effectively decouples the low-$z$ calibrators from the high-$z$ SNIa sample, thereby yielding an excellent $\Delta\chi^{2}$. However, an additional feature is visible in Fig.~\ref{fig:chi2}: both the Hstep+Mstep and the $w_{0}w_{a}$+Mstep models exhibit a noticeable bump around $\zt \sim 0.15$.

This feature can be understood from Fig.~\ref{fig:zt_0.01_or_0.15}. For the Hstep+Mstep model, a transition at $\zt = 0.013$ produces an evolution that is qualitatively similar to a pure Mstep model with a comparable $\zt$. In contrast, the same model with $\zt = 0.15$ behaves very differently. At higher transition redshifts, an $M$ step introduces a discontinuity which affects the high-$z$ SNIa and which is strongly disfavored. The accompanying $H$ transition compensates for this by shifting the expansion rate so as to keep the effective $\mathcal{M}$ approximately constant. However, this compensation can only operate up to a certain redshift: if the $H$ transition begins at a significantly higher redshift than $z\sim 0.15$, it leads to unacceptable deviations in the BAO distances. Consequently, there exists a ``sweet spot'' around $z \sim 0.15$ where a modest $M$ transition, combined with a mild modification to the expansion history, partially alleviates the tension and yields a small (though not dramatic) improvement, with $\Delta\chi^2\simeq -16$.

The presence of the aforementioned sweet spot in redshift is primarily driven by the DESI BGS measurement at a redshift of $z \sim 0.3$, which effectively constrains the range in which such a feature can occur. This result highlights the importance of future observations in the low-$z$ regime. In particular, an additional measurement in the interval $0.1 < z < 0.3$ would be especially valuable for discriminating and potentially falsifying this class of ``partial solutions.''

A similar mechanism operates in the $w_{0}w_{a}$+Mstep model and even in the non-minimally coupled scalar-field model. In the latter case, the effective $M$ transition occurs around the epoch when dark energy begins to dominate ($\zt \sim 1$). This naturally produces a relatively weak $M$ variation coupled with a highly flexible expansion history, leading to a moderate improvement in $\Delta\chi^{2}$.

An additional noteworthy feature of these models is their ability to accommodate a higher value of $H_0$, as shown in Table~\ref{tab:results2}. A larger $H_0$ partially alleviates the requirement for a large $M$ transition, which is reflected in the $M$ and ${M}_E$ columns of the table. Crucially, this increase in $H_0$ occurs only in models that allow a flexible expansion history. In a $\Lambda$CDM-like expansion history, a Hubble constant of $\sim 70$--$72\,\kmsMpc$ would lead to tension with BAO distance measurements. In contrast, in the Hstep+Mstep model this issue is mitigated by the presence of $\HnodE$; the value $\HnodE \simeq 68$ km\,s$^{-1}$\,Mpc$^{-1}$ maintains consistency with BAO distances despite the higher inferred $H_0$.

A similar mechanism operates in the $w_0 w_a +$Mstep model: although the model contains only a single $H_0$, the freedom introduced by the parameters $w_0$ and $w_a$ in the expansion history allows the BAO constraints to be satisfied with a moderately acceptable fit ($\dchisq=-16$ relative to \lcdm\ and $H_0\simeq 70\kmsMpc$). Finally, in the non-minimal coupling model, there is again a single $H_0$, but the expansion history gains additional flexibility through the three free parameters $\xi$, $\beta$, and $m$. As a result, the fit is even better ($\dchisq=-27$ relative to \lcdm) and this model can accommodate an even higher value of the Hubble constant ($H_0\simeq 72\kmsMpc$) than the other models; see again Table \ref{tab:results2}.

\bibliographystyle{apsrev4-2}
\bibliography{references}

\end{document}